\documentclass[12pt]{article}
\usepackage[latin9]{inputenc}
\usepackage{amsmath}
\usepackage{amssymb,color}
\usepackage{graphicx}
\usepackage{esint}
\usepackage{cite}
\makeatletter
\usepackage{subfigure}
\usepackage{amscd}
\usepackage{appendix}
\usepackage{verbatim}
\usepackage{amsmath,bm}

\begin{document}
	\begin{titlepage}		{\hbox to\hsize{\hfill{} CTPU-PTC-21-16 }}
		
		\bigskip{}
		\vspace{3\baselineskip}
		
		\begin{center}
			\textbf{\Large{Gravitational Leptogenesis in Bounce Cosmology}}\par
		\end{center}{\Large \par}
		
		\vspace{0.1cm}
		\begin{center}
			\textbf{
				Neil D. Barrie
			}\\
			\textbf{ }
			\par\end{center}
		
		\begin{center}
			{\it
			 		Center for Theoretical Physics of the Universe, Institute for Basic Science (IBS),
			 Daejeon, 34126, Korea\\ 
	Email: nlbarrie@ibs.re.kr}\\
			\textit{\small{}}
			\par\end{center}{\small \par}
		
		\begin{center}
			\textbf{\large{}Abstract}
			\par\end{center}{\large \par}
		
		\noindent 
We investigate whether successful Gravitational Leptogenesis can take place during an Ekpyrotic contraction phase. Two possible paths by which this can occur are coupling the  Ekpyrotic scalar to a gravitational Chern-Simons term, or to a  $ U(1) $ gauge field Chern-Simons term. These couplings lead to the production of chiral gravitational waves, which generate a lepton number asymmetry through the gravitational-lepton number anomaly. This lepton asymmetry is subsequently reprocessed by equilibrium sphaleron processes to produce a baryon asymmetry. We find successful Gravitational Leptogenesis to be possible in Ekpyrotic bounce cosmologies through both of these mechanisms.

	\end{titlepage}
	\section{Introduction}

One of the major mysteries of modern particle physics and cosmology is the source of the observed matter-antimatter asymmetry of the universe.  The size of this asymmetry is described by the quantity $\eta_{B}$ \cite{Aghanim:2018eyx}, known as the baryon asymmetry parameter,
\begin{equation}
\eta_B = \frac{n_B}{s} \simeq 8.5\times 10^{-11} ,
\label{eta_param}
\end{equation}
where $n_B$ and $s$ are the baryon number and entropy densities of the universe, respectively. To produce such an asymmetry in the early universe, three criteria   must be satisfied which are known as the Sakharov conditions\cite{Sakharov:1967dj}. Although the Standard Model (SM) is able to satisfy each of these requirements, the predicted asymmetry parameter is many orders of magnitude too small to explain observations \cite{Cohen:1993nk}. This result necessitates the existence of physics beyond the SM, and has motivated the investigation of many possible extensions. 

 One particularly intriguing proposal for generating the observed baryon asymmetry is the scenario known as Gravitational Leptogenesis. Within the SM, there exists a gravitational anomaly associated with the $ U(1) $ lepton number global symmetry, due to the absence of right-handed neutrinos \cite{AlvarezGaume:1983ig}. This provides a pathway for lepton number violation if a non-zero Chern-Pontryagin density ($ R\tilde{R} $) is generated in the early universe. One way to achieve this is the production of chiral gravitational waves through an interaction between the  gravitational  Chern-Simons term and a rolling pseudoscalar field, that is $\phi R\tilde{R}$ \cite{Lue:1998mq,Choi:1999zy}. If the pseudoscalar is identified as the inflaton, its motion during the inflationary epoch can lead to the production of high frequency chiral gravitational waves which in turn generate a non-zero lepton number density through the anomaly \cite{Alexander:2004us,Lyth:2005jf,Fischler:2007tj,Kawai:2017kqt,Adshead:2017znw,Kamada:2019ewe,Kamada:2020jaf,Samanta:2020tcl}.  After reheating, the lepton asymmetry is reprocessed by equilibrium sphaleron processes prior to the Electroweak Phase Transition (EWPT), as in usual Leptogenesis scenarios \cite{Kuzmin:1985mm,Fukugita:1986hr}. Unfortunately, this simple application of this mechanism is unable to produce a large enough asymmetry to explain observations while remaining consistent with current inflationary constraints, and the required absence of graviton ghost modes \cite{Alexander:2004us,Lyth:2005jf,Fischler:2007tj}. Although, in recent work it has been shown that with the inclusion of a period of kination prior to reheating, the standard inflationary Gravitational Leptogenesis scenario can successfully explain observation \cite{Kamada:2019ewe}. 
 
 To generate a non-zero $ R\tilde{R} $ it is not necessary for the pseudoscalar to couple to the gravitational Chern-Simons term, but rather through a gauge field Chern-Simons interaction of the form  $ \varphi F\tilde{F} $. This interaction can lead to the significant production of helical gauge fields as $ \varphi $ rolls, which subsequently generate  chiral gravitational waves \cite{Cook:2011hg,Ben-Dayan:2016iks}. A key advantage of this form of Gravitational Leptogenesis is that it avoids the possible issues of graviton ghost modes encountered by the introduction of the gravitational Chern-Simons term. In investigating this scenario, it was found that successful inflationary Gravitational Leptogenesis was possible to achieve when considering non-abelian $ SU(2) $ gauge fields \cite{Noorbala:2012fh,Maleknejad:2014wsa,Maleknejad:2016dci,Caldwell:2017chz}, but not for an abelian $ U(1) $ gauge field \cite{Papageorgiou:2017yup}.

 Inflationary  Cosmology provides a unique venue for Baryogenesis alongside its many successes at explaining cosmological observables \cite{ Alexander:2011hz,Barrie:2014waa, Barrie:2015axa,Anber:2015yca,Jimenez:2017cdr,Domcke:2019mnd}. The past explorations of Gravitational Leptogenesis have focused on the inflationary setting for production of chiral gravitational waves. The inflationary scenario still suffers from many unresolved problems, motivating the investigation of  alternative cosmological models \cite{Novello:2008ra,Lehners:2008vx,Brandenberger:2009jq,Battefeld:2014uga,Brandenberger:2016vhg,Ijjas:2018qbo}. One important example of this is Bounce Cosmologies, and in particular the Ekpyrotic universe \cite{Khoury:2001wf,Lyth:2001pf,Brandenberger:2001bs,Khoury:2001zk,Lyth:2001nv,Steinhardt:2001st,Khoury:2003rt,Lehners:2007ac,Buchbinder:2007ad,Buchbinder:2007tw,Cai:2012va,Ijjas:2019pyf}. The main idea of a bounce cosmology is that there existed a period of space-time contraction prior to the onset of standard Big Bang Cosmology. These two epochs are separated by a bounce through which the universe transitions from a period of contraction to the usual expansion phase. The  well-studied Ekpyrotic Cosmology involves a period of ultra-slow contraction ($ \omega\gg 1 $) prior to a bounce. This type of contraction phase can  be induced by a fast-rolling scalar field that has a negative exponential potential. The large equation of state means that in a contracting universe, the scalar field will quickly come to dominate the energy density of the universe beginning an Ekpyrotic contraction phase. Thus, resulting in the dilution of any anisotropies and other initial energy densities. Much like inflationary cosmology, the  Ekpyrotic scenario solves each of the flatness, horizon and monopole problems, and is capable of generating the perturbations observed in the Cosmic Microwave Background (CMB). The simplest form of the Ekpyrotic universe, induced by a single scalar field, is  in tension with current observational results, but many theoretical developments and improvements have been made since this initial proposal \cite{Battefeld:2014uga, Buchbinder:2007ad,Buchbinder:2007tw}. In analogy to inflationary cosmology, Ekpyrotic models can have many intriguing phenomenological implications. Recent studies have investigated its applications to various cosmological phenomena -   Baryogenesis, the origin  of dark matter, Magnetogenesis and gravitational waves \cite{Salim:2006nw,Li:2014era,Cheung:2014nxi,Ben-Dayan:2016iks,Ito:2016fqp,Koley:2016jdw,Barrie:2020kpt}. Periods of Ekpyrotic contraction  are common features of many cyclic and bounce cosmologies, so it is interesting to consider the possible phenomenological consequences of such an epoch.

In this work, we explore the possibility of successful Gravitational Leptogenesis taking place during an Ekpyrotic contraction phase in a Bounce Cosmology. This work is inspired by the connection between  Ekpyrotic Cosmology and Baryogenesis that was first drawn in Ref. \cite{Barrie:2020kpt}. The  blue-shifted gravitational wave predictions associated with an epoch of Ekpyrotic contraction  may alleviate the stringent CMB constraints that strongly constrain the inflationary versions of Gravitational Leptogenesis. Two paths for successful Gravitational Leptogenesis will be considered in this work, the coupling of the Ekpyrotic scalar to the gravitational Chern-Simons term, and to an abelian  $ U(1) $ gauge field Chern-Simons term. The paper will be structured as follows; , Section \ref{EKP} will provide a brief discussion of the properties and observational features of the Ekpyrotic phase. In Section \ref{Grav_lep_path1}, the gravitational Chern-Simons coupling scenario  will be explored, and the phenomenological constraints and allowed parameter space determined. Section \ref{Grav_lep_path2} investigates the chiral gauge field production scenario for Gravitational Leptogenesis and the requirements for successful Baryogenesis, along with the potential for additional high frequency gravitational wave observables. Finally, in Section \ref{conc}, we will conclude with a discussion of the implications of the results and future directions for investigation.

\section{Ekpyrotic Cosmology}
\label{EKP}

The Inflationary scenario is a well-established paradigm in standard cosmology due to its success at solving various observational problems such as the flatness, horizon, and monopole problems, as well as providing measurable predictions in the form of primordial perturbations \cite{Guth:1980zm,Linde:1981mu,Albrecht:1982wi,Mukhanov:1981xt}. Many models have been proposed and significant effort expended in the pursuit of experimental verification, but the exact mechanism for inflation is unclear \cite{Martin:2013tda}. These include questions surrounding initial conditions, fine-tuning,  the singularity problem, degeneracy of model predictions, trans-Planckian field values and violation of perturbativity \cite{Brandenberger:2016vhg}.  The known issues with inflation have led to the exploration of possible alternatives to the usual inflationary paradigm, such as  string gas cosmology, bounce, and cyclic models. As with inflation, these models attempt to solve the flatness, horizon, and monopole problems, and must be able to source the nearly scale invariant spectrum of temperature fluctuates observed in the CMB. In what follows, we will focus on a well-known type of bounce cosmology, the Ekpyrotic bounce, which solves the known cosmological problems, and can potentially resolve various issues with other bounce models, while providing the benefits over inflation of geodesic completion, sub-Planckian field values, and removal of the singularity problem. This type of contraction phase is a  feature of many bounce cosmologies, such as recent studies into cyclic universe models \cite{Ijjas:2019pyf}.

An Ekpyrotic period of contraction is characterised by a large equation of state $ \omega\gg 1 $ . This cosmological evolution can be induced by a stiff form of matter such as a fast-rolling scalar field. In a contracting universe, the stiff matter will come to dominate the total energy density of the universe,
\begin{equation}
\rho_{\textrm{Total}}=\frac{\rho_k}{a^2}+\frac{\rho_{\textrm{mat}}}{a^3}+\frac{\rho_{rad}}{a^4}+\frac{\rho_{a}}{a^6}+...+\frac{\rho_\varphi}{a^{3(1+\omega_\varphi)}}+...~,
\label{ener_den}
\end{equation}
where $ \rho_a $ is the energy density associated with anisotropies, and $ \rho_\varphi $ is the energy density of the fields responsible for the Ekpyrotic contraction. From Eq. (\ref{ener_den}) it is clear that in a contracting space-time background the $ \rho_\varphi $ term will quickly increase relative to the other energy densities, thus dominating the energy density of the universe, for $ \omega_\varphi\gg 1 $. Consequently, a sufficiently long period of $ \omega_\varphi\gg 1 $ contraction naturally leads to the suppression of any curvature and anisotropy perturbations, while also diluting the initial radiation and matter densities. This is how the Ekpyrotic phase solves the known cosmological problems, while also remedying the problems of  initial anisotropy growth and anisotropic instabilities that can occur in other bounce scenarios.

To see how  an Ekpyrotic contracting epoch can be induced by a rolling scalar field, consider the following  relation for the equation of state parameter for a scalar $ \varphi $,
\begin{equation}
\omega=\frac{\frac{1}{2}\dot{\varphi}^2 -V(\varphi)}{\frac{1}{2}\dot{\varphi}^2 +V(\varphi)} ~.
\end{equation}
An equation of state of $ \omega\gg 1 $ can be achieved if,
\begin{equation}
 \frac{1}{2}\dot{\varphi}^2+V(\varphi) \approx 0   ~~\text{and}~~  \frac{1}{2}\dot{\varphi}^2 - V(\varphi) \gtrsim 0 ~.
 \end{equation}
 
 One way to satisfy these relations is for the scalar $ \varphi $ to be fast-rolling down a negative exponential potential, leading to an approximate cancellation in the denominator. That is, a scalar potential of the following form,
\begin{equation}
V(\varphi)\approx-V_0 e^{-\sqrt{2\epsilon}\frac{\varphi}{M_p}} ~,
\end{equation}
where the $ \epsilon $ parameter shall be referred to as the fast-roll parameter, and $M_p=2.4\cdot 10^{18}$ GeV is the reduced Planck mass \cite{Baumann:2014nda}. The relation between $ \epsilon $ and $ \omega $ is given by,
\begin{equation}
\epsilon=\frac{3}{2}(1+\omega) ~.
\end{equation} 
 
The fast-roll parameter can be considered analogous to the inflationary slow-roll parameter where $ \epsilon_{\textrm{inf}}\ll 1 $, while $ \epsilon \gg 1 $ and $ \varphi $ satisfies corresponding fast-roll conditions \cite{Khoury:2003rt}. Interestingly, there is a seeming duality between the Ekpyrotic and Inflationary regimes through the respective fast and slow-roll parameters, which was the initial motivation for our previous work \cite{Barrie:2020kpt}, and the model considered here.

The Ekpyrotic action is of the following form,
\begin{equation}
S_\textrm{Ekp}=\int d^4 x \sqrt{-g}\left(\frac{M_p^2}{2} R -\frac{1}{2}\partial_\mu \varphi\partial^\mu \varphi +V_0  e^{-\sqrt{2\epsilon}\frac{|\varphi|}{M_p}}\right)~,
\label{ekp_scalar}
\end{equation}
from which we derive the scale factor describing the Ekpyrotic contraction epoch, 
\begin{equation}
a= (\epsilon H_b t)^\frac{1}{\epsilon} =(\epsilon H_b  \tau)^\frac{1}{\epsilon-1}~,
\label{sca_fac} 
\end{equation}
and corresponding Hubble rate,
\begin{equation}
H= \frac{H_b}{ (\epsilon H_b|\tau|)^{\frac{\epsilon}{\epsilon-1}}} \simeq\frac{1}{\epsilon \tau}~,
\label{Hub}
\end{equation}
where we fix the bounce point to be at $ t_b=\tau_b= \frac{1}{\epsilon H_b} $ such that $ a(\tau_b)=a(t_b)=1 $, and $ t,\tau\in(-\infty,\frac{1}{\epsilon H_b} ) $ during the Ekpyrotic contraction. For large $ \epsilon $, the scale factor evolution is very slow such that $ t\sim \tau $. 
Through inspection of Eq. (\ref{sca_fac}) and Eq. (\ref{Hub}), it is clear that for $ \epsilon\gg 1 $ the Hubble rate can increase exponentially, while the scale factor shrinks by only an $ \mathcal{O}(1) $ factor. Thus, for $ \epsilon \sim \mathcal{O}(100) $ only a single e-fold of contraction is required to generate 60 e-folds worth of perturbations.

In the case of $ \epsilon\gg 1 $, the equations of motion for the  scalar $ \varphi $ are solved by the scaling solution \cite{Buchbinder:2007ad},
\begin{equation}
\varphi \simeq M_p \sqrt{\frac{2}{\epsilon}} \ln(-\sqrt{\epsilon V_0}\tau/M_p)~,
\label{phi_sol}
\end{equation}
and subsequently,
\begin{equation}
	\varphi^\prime\simeq\sqrt{\frac{2}{\epsilon}}\frac{M_p}{\tau}~,
	\label{scaling}
\end{equation}
which are expressed in conformal time.

Despite the advantages and simplicity of the scenario described above, the single field Ekpyrotic scenario leads to a blue-tilted spectral index, which is in significant tension with current CMB observations \cite{Battefeld:2014uga,Aghanim:2018eyx}.  This is one of the main issues of the original formulation of Ekpyrotic Cosmology. One way to alleviate this issue is the introduction of an additional Ekpyrotic scalar field. If the Ekpyrotic contraction is followed by a period of kinetic dominated contraction ($ \omega=1 $) prior to the bounce, the nearly scale invariant scalar power spectrum in the CMB can be produced for $ \epsilon \sim \mathcal{O}(100) $ through the conversion of isocurvature perturbations into adiabatic perturbations by the additional scalar \cite{Ijjas:2015hcc,Levy:2015awa}. This is known as the New Ekpyrotic model, in which the background evolution is induced by two Ekpyrotic scalars and consists of a non-singular bounce sourced by a ghost condensate \cite{Buchbinder:2007ad,Lehners:2007ac,Kallosh:2007ad,Buchbinder:2007tw,Cai:2012va,Ijjas:2019pyf}. In this work we will mainly focus on the simplest single field form of the Ekpyrotic scenario, with the possibility of embedding it into Bounce Cosmology models which contain an epoch of Ekpyrotic contraction.

The New Ekpyrotic model tends to predict relatively large non-gaussianities in the CMB, compared to inflationary models. This can provide constraints on the background evolution in combination with the scalar power spectrum. The current best constraints on the non-gaussianities, from the Planck observations \cite{Akrami:2019izv}, are,
\begin{equation}
f_{NL}^{\textrm{local}} =-0.9\pm 5.1 ~,
\label{NG}
\end{equation} 
while the $ f_{NL} $ predicted by the two scalar Ekpyrotic scenario, with a period of kination prior to the bounce,  is \cite{Buchbinder:2007at,Lehners:2008my,Lehners:2010fy},
\begin{equation}
f_{NL}\propto \sqrt{\epsilon}~,
\label{NG_ekp}
\end{equation} 
where $ \epsilon \sim \mathcal{O}(100) $ can successfully produce  a nearly scale-invariant scalar power spectrum. The general form of the kinetic conversion scenario is in some tension with current observations, but can be resolved via modifications to the scalar sector. In the models presented in Ref. \cite{Fertig:2013kwa,Ijjas:2014fja}, zero non-gaussianities are generated during the Ekpyrotic contraction phase, instead they are only produced during the conversion process prior to the bounce. This reduces the non-gaussianities to $ f_{NL}\sim \mathcal{O}(1) $ with dependence on the form of interactions between the two scalars, and efficiency of the conversion process. Thus, increased  precision in measurements of the non-gaussianities alongside improvements in the theoretical understanding of the period around the bounce point and the Ekpyrotic scalar sector are necessary.

 Another characteristic of Ekpyrotic Cosmologies is that they predict a blue-tilted tensor power spectrum with a small tensor-to-scalar ratio $ r $ on CMB scales, that is below current sensitivities and difficult to measure within the near future \cite{Boyle:2003km}. The tensor perturbation spectrum is given by,
\begin{equation}
 \mathcal{P}^{v}_{T}\simeq \frac{4 k^2}{\pi^2 M_p^2}~,
 \label{Ekp_GW}
\end{equation}
where $ \epsilon \gg 1 $ has been assumed. Thus, if near future experiments such as LiteBIRD \cite{Hazumi:2019lys} are able to observe a  tensor-to-scalar ratio, significant constraints will be applied on the standard Ekpyrotic scenario. It was recently shown that the period around the bounce can alter the predicted gravitational wave spectrum, leading to a larger prediction of $ r $ at CMB scales \cite{Brandenberger:2020tcr}. In the Gravitational Leptogenesis scenario considered in Section \ref{Grav_lep_path2},  the gravitational waves produced by the enhanced production of gauge fields, may lead to additional high frequency gravitational wave signatures \cite{Ben-Dayan:2016iks, Ito:2016fqp}.

A common  feature of Ekpyrotic scenarios is a period of kinetic expansion after the bounce, such as in the Matter-Ekpyrotic Bounce \cite{Cai:2012va}. A period of kination could enhance the generated lepton asymmetry because the energy stored in the rolling scalar field, and associated maximum reheating temperature, decreases faster than the matter energy density of the asymmetry dilutes. This has been utilised in the inflationary Gravitational Leptogenesis scenarios to increase the predicted baryon asymmetry \cite{Papageorgiou:2017yup,Kamada:2019ewe}, and will be included as part of our analysis.

\section{Gravitational Leptogenesis via $ f(\varphi) R \tilde{R} $}
\label{Grav_lep_path1}
Firstly, we will investigate the standard Gravitational Leptogenesis within the context of a  period of Ekpyrotic contraction.  The SM contains a gravitational-$ U(1) $ anomaly associated with the global lepton number symmetry, due to the absence of right-handed neutrinos. The gravitational lepton anomaly has been found to have many interesting phenomenological effects in the early universe, such as possible birefringent propagation of gravitational waves \cite{Barrie:2017mmr}.  Gravitational Leptogenesis utilises the existence of the gravitational-lepton anomaly within the SM, through which a chiral asymmetry in the gravitational sector can be transmitted to the lepton sector. The gravitational anomaly and its relation to the lepton current is given by,
\begin{equation}
	\nabla_{\mu}J_L^{\mu} = \frac{N_{\mathrm{R-L}}}{384\pi^2}R\tilde{R},  \label{grav_anom}
\end{equation}
where the Chern-Pontryagin density is defined as,
\begin{equation}
R\tilde{R} \equiv R_{\mu\nu\alpha\beta}\tilde{R}^{\mu\nu\alpha\beta} = \frac{1}{2}\frac{\epsilon^{\alpha\beta\gamma\delta}}{\sqrt{-g}}R_{\alpha\beta\rho\sigma}R_{\gamma\delta\mu\nu}g^{\mu\rho}g^{\nu\sigma}~,
\end{equation}
and $N_\mathrm{R-L}$ is the difference between the number of left and right-handed leptons in the model. In what follows we will assume that this takes the value of $N_\mathrm{R-L}=3$, as in the SM where right-handed neutrinos are absent. Even with the inclusion of right-handed neutrinos the production of a lepton asymmetry is still possible below the typically high mass scale of the right-handed neutrinos. Their masses would then constitute an effective cut-off scale, which may be as high as $ 10^{16} $ GeV. These high scale right-handed neutrinos have the added benefit of being able to explain the observed tiny neutrino masses through the seesaw mechanism \cite{Yanagida:1979as,GellMann:1980vs}. Although masses larger than $ 10^{16} $ GeV  require the introduction of non-perturbative neutrino Yukawa couplings, and subsequently will not be considered here. Interestingly, Gravitational Leptogenesis is still possible  in the case of Dirac neutrinos, despite the cancellation of the gravitaional anomaly \cite{Adshead:2017znw}. 


In this section, we consider the source of the chiral gravitational waves and subsequent lepton asymmetry to be an interaction between the Ekpyrotic scalar, $ \varphi $, and the gravitational Chern-Simons term. This coupling is of the following form,
\begin{equation}
f(\varphi)R\tilde{R}~,
\end{equation}
where an interaction term of this type could originate from the Green-Schwarz mechanism, and may arise within the context of string theory \cite{Green:1984sg}. The dynamics of the gravitational Chern-Simons coupling is dictated by the motion of the pseudoscalar field $ \varphi $ in its potential, with associated production of parity violating gravitational waves.  In what follows, we will define the function $ f(\varphi) $ as,
\begin{equation}
f(\varphi)=\frac{\mathcal{N}}{16 \pi^2 M_p^3}\varphi~,
\label{coupling}
\end{equation}
where $ \mathcal{N} $ defines the strength of the Chern-Simons coupling, and may be related to the string theory compactification. In the inflationary setting, the size of this coupling is constrained by the cosmological observables and the required non-existence of graviton ghost modes.

In the inflationary version of Gravitational Leptogenesis, it is not possible to obtain the observed baryon asymmetry unless there is a period of kination prior to the onset  of reheating. For the purposes of comparison to the inflationary case, we will consider the results found for standard Gravitational Leptogenesis in Ref. \cite{Kamada:2019ewe}, and utilise the notation therein. Much of the approach for the inflationary case is translatable to the Ekpyrotic case, by substituting for the scale factor and the alternative time dependence of the gravitational Chern-Simons coupling. These two cosmologies also exhibit a duality in replacing the slow roll parameter with the inverse of the fast roll parameter.

 To begin we introduce the  important terms of the action,
\begin{equation}
	S = \int d^4x\sqrt{-g}\left(\frac{M_p^2}{2}R -\frac{1}{2}\partial_\mu \varphi\partial^\mu \varphi +V_0  e^{-\sqrt{2\epsilon}\frac{|\varphi|}{M_p}} +\frac{M_p^2}{4}f(\varphi)R\tilde{R}\right)~, \label{fullS}
\end{equation}
including the Einstein-Hilbert term and the $ \varphi $ dependent terms.

To determine the lepton number density generated during the contraction phase, we must derive the equations of motion for $ \varphi $ and the gravitational tensor perturbation $ h_{ij} $. We will assume that the background evolution is dominated by the Ekpyrotic scalar and that the gravitational Chern-Simons term has a negligible impact on its evolution. This gives us the scaling solution presented in the previous section, namely,
\begin{equation}
\varphi^\prime\simeq\sqrt{\frac{2}{\epsilon}}\frac{M_p}{\tau}~,
\label{scaling1}
\end{equation}
where we have assumed the sign is such that the resultant baryon asymmetry will be in the matter rather than antimatter sector.

The gravitational wave dynamics can be calculated by considering the following perturbed FRW metric,
\begin{equation}
ds^2 = a(\tau)^2(-d\tau^2 + (\delta_{ij} + h_{ij})dx_idx_j)~,
\label{metr}
\end{equation}
where we take the transverse-traceless gauge for $h_{ij}$, and the scale factor is defined as $ a=(\epsilon H_b  \tau)^\frac{1}{\epsilon-1} $ . Substituting this into the action in Eq. (\ref{fullS}), we obtain,
\begin{align}
	S = \frac{M_p^2}{8}\int d^4x & \left( a^2(\tau)((h^i_{\ j})'(h^j_{\ i})' - (\partial_kh^i_{\ j})(\partial^kh^j_{\ i}))\right.\nonumber \\ 
	& \left.- f'(\varphi)\epsilon^{ijk}((h^l_{\ i})'(\partial_jh_{kl})' - (\partial^mh^l_{\ i})\partial_j\partial_mh_{kl}) \right)+...~, 
	\label{grav_ac}
\end{align}
 where the $ ... $ represents the $ \varphi $ kinetic term and potential.
 
 To calculate the generated $ R\tilde{R} $, we must consider the asymmetry of the left and right-handed circular polarisations of the tensor perturbations. To do so, we must first decompose the tensor perturbations into the circular polarisations as follows,
\begin{equation}
h_{ij}(\tau,\vec x) = \frac{1}{(2\pi)^{3/2}}\int d^3k\sum_{q = \mathrm{R,L}}p_{ij}^q(\vec k)h_q(\tau)e^{i\vec k\cdot \textbf{x}}~, \label{circ_pol}
\end{equation}
where we can relate the circular ($p_{ij}^{R, L}(\vec k)$) and linear  ($p_{ij}^{+, \times}(\vec k)$) polarization tensors  by, 
\begin{align}
	p_{ij}^\mathrm{R} &= (p^+_{ij} + ip_{ij}^{\times})/\sqrt{2}~, \nonumber \\
	p_{ij}^\mathrm{L} &= (p^+_{ij} - ip_{ij}^{\times})/\sqrt{2} = (p^R_{ij})^*~, 
\end{align}
in which the linear polarization tensors are given by,
\begin{align}
	p_{ij}^+  &= (\varepsilon_1)_i  (\varepsilon_1)_j - (\varepsilon_2)_i  (\varepsilon_2)_j ~,  \nonumber \\
	p_{ij}^\times  &= (\varepsilon_1)_i  (\varepsilon_2)_j + (\varepsilon_1)_j  (\bm{\varepsilon}_2)_i~, 
\end{align}              
and we have defined the right-handed, orthogonal triad of unit vectors $(\varepsilon_1({\vec k}), \varepsilon_2({\vec k}), \varepsilon_3({\vec k}))$ with  $|{\vec k}|\varepsilon_3({\vec k}) = \vec k$ and  $\varepsilon_{1,2}(-{\vec k})=-\varepsilon_{1,2}(-{\vec k}) $. This then gives us the following relations for the circular polarization tensors,
\begin{align}
	p_{ij}^\mathrm{R}p^{ij\mathrm{R}} &= p^\mathrm{L}_{ij}p^{ij\mathrm{L}} = 0~,\nonumber \\
	p_{ij}^\mathrm{R}p^{ij\mathrm{L}} &= 2~,\nonumber \\
	k_m\epsilon^{lmj} p^{\mathrm{R,L}}_{ij}  &= -i\lambda^{\mathrm{R,L}} k \ p^{l \  \mathrm{R,L}}_{\ \ i}~, 
\end{align}
where we have defined $\lambda^\mathrm{R} = +1, \lambda^\mathrm{L} = -1$ and suppressed the $ \vec k $ dependencies. 
Now we can write the action in terms of the circular polarisations, using the above relations and the action components given in Eq. (\ref{grav_ac}),
\begin{equation}
S = \frac{M_p^2}{4} \int d\tau d^3k \sum_{\lambda = \pm} a^2(\tau)\left(1 - \lambda k\frac{f'(\tau)}{a^2(\tau)}\right)(|(h_\lambda)'|^2 - k^2|h_\lambda|^2)+...  \label{S2_f}
\end{equation}
To derive the wave mode function equations we make the following reparametrisation,
\begin{equation}
\mu_\lambda = z_\lambda h_\lambda~,
\end{equation}
where
\begin{equation}
z_\lambda^2 =\frac{a(\tau)^2 M_p^2 }{2} \left( 1 -  \lambda k\frac{f'(\tau) }{a(\tau)^2M_p}\right)~,
\end{equation}

  In this parametrisation the following wave mode function equation for $ \mu_\lambda $ can be derived,
\begin{equation}
\mu_\lambda^{\prime \prime} +\left(k^2 - \frac{z_{\lambda}^{\prime\prime}}{z_{\lambda}}\right)\mu_\lambda =0~.
\end{equation}

Up to this point, our analysis of the tensor perturbations is the same as that presented in the inflationary scenario. When we now specify $ \varphi $ as the Ekpyrotic scalar, assume the induced background evolution is that of Ekpyrotic contraction, and substitute Eq. (\ref{coupling}) we obtain,
\begin{equation}
z_\lambda^2 =\frac{(\epsilon H_b \tau)^{2/(\epsilon-1)}M_p^2 }{2 } \left( 1- \lambda C \frac{k}{\tau (\epsilon H_b \tau)^{2/(\epsilon-1)}}\right)\simeq \frac{M_p^2 }{2 } \left( 1-\lambda C \frac{k}{\tau}\right) ~,
\label{z_term}
\end{equation}
where
\begin{equation}
C= \frac{\mathcal{N}}{16 \sqrt{2\epsilon} \pi^2 M_p^2}~,
\end{equation}
and $ \epsilon\gg 1 $ has been assumed, for which the scale factor can be taken to be approximately constant. Thus, the wave mode equation for $ \mu_\lambda $ becomes approximately,
\begin{equation}
\mu_\lambda^{\prime \prime} +\left(k^2 +\frac{2}{\epsilon\tau^2}+ \lambda C\frac{k}{\tau^3}\right)\mu_\lambda =0~,
\label{Eommu}
\end{equation}
in the limit  $ \epsilon\gg 1 $.

The equation of motion in Eq. (\ref{Eommu}) does not have a closed form, so to proceed we will utilise the Wentzel-Kramers-Brillouin (WKB) approximation, which has also been utilised in other Gravitational Leptogenesis  and chiral gravitational wave studies \cite{Namba:2015gja,Kawai:2017kqt}. Thus, we obtain the following solution for $ \mu_{\lambda} $,
\begin{equation}
\mu_\lambda \simeq  \frac{1}{\sqrt{2 k} \left(1 +\frac{2}{\epsilon (k\tau)^2}+ \lambda\frac{C}{k \tau^3 }\right)^{\frac{1}{4}}} e^{-i k\int\sqrt{1 +\frac{2}{\epsilon (k\tau)^2}+ \lambda\frac{C}{k \tau^3 }}d\tau }~,
\end{equation}
and the associated form for the perturbation $ h_{\lambda} $,
\begin{equation}
h_\lambda \simeq \frac{1}{M_p \sqrt{ k} \left(1+\frac{2}{\epsilon (k\tau)^2}+  \lambda\frac{C}{k \tau^3 }\right)^{\frac{1}{4}}\sqrt{1-\lambda\frac{kC}{ \tau }}} e^{-i k\int\sqrt{1 +\frac{2}{\epsilon (k\tau)^2}+ \lambda\frac{C}{k \tau^3 }}d\tau } \label{WKB}~,
\end{equation}
where we have matched to the initial plane wave solution,
\begin{equation}
\mu_\lambda = \frac{M_p}{\sqrt{2}} h_\lambda  \simeq \frac{1}{\sqrt{2 k}} e^{-i k\tau }~,
\end{equation}

To ensure that ghost modes do not appear we must require that $1> \left|\frac{C k}{\tau}\right|$ for the entire Ekpyrotic phase, which gives the following constraint on the model parameters, 
\begin{equation}
10^7 >\mathcal{N} \left(\frac{\epsilon}{5 \cdot 10^4}\right)^{1/2} \left(\frac{\Lambda_e}{10^{16} \textrm{~GeV}}\right)^{-1}\left(\frac{H_b}{2 \cdot 10^{11} \textrm{~GeV}}\right)~,
\label{WKB_con}
\end{equation}
where  we have assumed $ \Lambda_e $ is the UV cut-off of $ k $. If we assume for simplicity that $ \Lambda_e=\epsilon H_b=10^{16}  $ GeV, given by the maximal right-handed neutrino mass, we find the following constraint,
\begin{equation}
10^5> \mathcal{N}/\sqrt{\epsilon}~.
\label{WKB_con1}
\end{equation}

Interestingly, the Ekpyrotic case appears to not suffer the inability to define a consistent Bunch-Davies vacuum as seen in the inflationary scenario \cite{Lyth:2005jf}. Consider the large $ |\tau| $ limit in Eq. (\ref{z_term}), in the very early stage of Ekpyrotic contraction, $\tau \rightarrow -\infty$, the $ \mathcal{CP} $ violation term vanishes. This is in contrast to the inflationary case, in which this term has a linear dependence on $ \tau $ and thus can become negative at early times. We must also require that graviton ghost modes do not appear across the relevant range of $ k $ and $ \tau $. This requirement which is given in Eq. $ (\ref{WKB_con}) $ simultaneously ensures consistency of the WKB approximation.

Now that we know the dynamics of the tensor perturbations during the contraction phase, we can begin the calculation of the generated lepton number density.  From Eq. (\ref{grav_anom}), we see that in order to generate a lepton number asymmetry density, the Chern-Pontryagin density $ R\tilde{R} $ must have a non-zero expectation value. To determine the lepton number density we make the identification $n_L = a(\tau) J_L^0$, and expand the $R\tilde{R}$ term for the metric defined in Eq. (\ref{metr}). Upon taking the expectation value we arrive at the resultant lepton number density at a given time $ \tau $, in terms of the generated circularly polarised tensor perturbations,
\begin{equation}
n_L(\tau)= \frac{N_{R-L}}{768 \pi^4} \int^{\Lambda_e}_0 dk\left( k^5 \left( |h_+|^2-|h_-|^2 \right)-k^3 \left( | h_+|^2-|h_-^{\prime}|^2 \right)\right)~,
\label{nLden}
\end{equation}
where we have defined the UV momenta cut-off as $ \lambda_e $, and have assumed the scale factor is approximately constant across the Ekpyrotic phase for $ \epsilon \gg 1 $. In this derivation it has also been assumed that the initial Chern-Pontryagin density and $ n_L $ at the beginning of the Ekpyrotic phase are zero.

In evaluating this result, we are interested in the large $ k $ modes, which shall have the largest contribution to the lepton asymmetry. Upon evaluating Eq. (\ref{nLden}) for the WKB solution given in Eq. (\ref{WKB}) we arrive at the generated lepton number density at time $ \tau $,
\begin{equation}
n_L(\tau)= \frac{N_{R-L}}{1536 \pi^4}  \frac{\mathcal{N}}{16 \sqrt{2\epsilon} \pi^2 \Lambda M_p^3 |\tau|^3}\Lambda_e^4~,
\end{equation}
which evaluated at the bounce point $ \tau_b $ becomes, 
\begin{equation}
n_L(\tau_b)= \frac{1}{2048 \pi^6 } \frac{\mathcal{N} }{4\sqrt{2\epsilon}}\left(\frac{N_{R-L}}{3}\right)\left(\frac{\Lambda_e}{M_p}\right)^4 (\epsilon |H_b|)^3~,
\end{equation}
this result is of a similar form to that found in the inflationary scenario, but with the added enhancement from the fast roll parameter.

The lepton number density generated during the Ekpyrotic phase is assumed to be unchanged across the bounce point, and is matched to the end of the reheating epoch, which is taken as instantaneous and characterised by temperature $ T_{\rm rh} $. The exact dynamics of the bounce are model-dependent, and  are expected to  have only a minor effect as the bounce is assumed to be smooth and entropy conserving.  Hence the baryon number  density is given by,
\begin{equation}
n_B(\tau_b)=\frac{28}{79}n_L(\tau_b)\simeq 3.2\cdot 10^{-8} \frac{\mathcal{N} }{\sqrt{\epsilon}}\left(\frac{N_{R-L}}{3}\right)\left(\frac{\Lambda_e}{M_p}\right)^4 (\epsilon |H_b|)^3~,
\end{equation}
where we have assumed usual sphaleron redistribution of the net lepton asymmetry density prior to the EWPT. Using this relation, we can now calculate the predicted baryon asymmetry parameter, under the assumption of no additional entropy production after reheating; $ s=\frac{2 \pi^2}{45} g_* T_{\textrm{rh}}^3 $. The ratio of the predicted to observed asymmetry parameter is given by, 
\begin{equation}
\frac{\eta_B}{\eta_B^{obs}}\simeq \left(\frac{N_{R-L}}{3}\right) \left(\frac{\mathcal{N}}{ 6 \cdot 10^6}\right)  \left(\frac{\Lambda_e}{10^{16}~\textrm{GeV}}\right)^4 \left(\frac{ |H_b|}{ 2\cdot 10^{11}~\textrm{GeV}}\right)^{3/2} \left(\frac{\epsilon}{ 5\cdot 10^4}\right)^{5/2} ~,
\label{asym}
\end{equation}
where we have chosen reference parameters such that we maintain the non-existence of ghost modes and the consistency of the WKB approximation, and assuming instantaneous reheating $ H_{\textrm{b}}=\sqrt{11.7}\frac{T_{\textrm{rh}}^2}{M_p} $ . Imposing the constraint from Eq. (\ref{WKB_con1}), taking $ N_{R-L} =3$ and assuming $\Lambda_e= \epsilon H_b=10^{16}$ GeV, we find the following upper bound on the baryon asymmetry parameter ratio,
\begin{equation}
\frac{\eta_{B}}{\eta_{B}^{obs}}< \left(\frac{\epsilon}{2\cdot 10^4}\right)^{3/2}~,
\end{equation}
which corresponds to $ H_b<5 \cdot 10^{11} $ GeV, and a reheating temperature of $ T_{\textrm{rh}}< 6 \cdot 10^{14} $ GeV.

 Interestingly, we find that the instantaneous reheating scenario may lead to successful Gravitational Leptogenesis, in contrast to the inflationary alternative, which gives a maximum $ \frac{\eta^{\textrm{inf}}_{B}}{\eta_{B}^{obs}}\sim 2\cdot 10^{-9} $. The importance of the fast roll parameter in the difference between these predictions is encapsulated in this upper bound, and in Eq.  (\ref{asym})  where the $ \epsilon $ factor alone gives a $ 10^{10} $ contribution. Although it is possible to achieve the observed baryon number asymmetry, this requires model parameters that approach the upper limit of the validity of the wave mode function solution. One caveat in this result is that the successful Gravitational Leptogenesis requires a very large gravitational Chern-Simons coupling, which approaches the breakdown of the WKB approximation and possible existence of ghost modes. 
 
  In this analysis, we have ignored the potential washout effects that can occur due to thermal generation of the heavy Majorana neutrinos after inflation. Larger values of $ \epsilon $ allow for smaller $ H_b $ and thus reheating temperatures, but this may not allow the avoidance of washout effects. The washout factor for high reheating temperatures is expected to be of order $ 0.1 $ \cite{Adshead:2017znw}, despite which successful Gravitational Leptogenesis in this scenario can still be achieved.

In the previous result we assumed instantaneous and efficient reheating, but it is possible that the reheating epoch could be extended and characterised by inefficient entropy production. In this case, it could be possible to enhance the baryon asymmetry to entropy ratio, in turn opening up regions of the parameter space that were previously unfavourable. A period of kination after the bounce can lead to an enhancement of the asymmetry parameter \cite{Spokoiny:1993kt,Joyce:1996cp}. This epoch would be characterised by an equation of state $w=1$, and is a common feature of bounce cosmologies containing periods of Ekpyrotic contraction phases, as discussed in Section \ref{EKP}. This enhancement was utilised to save the inflationary Gravitational Leptogenesis scenario in Ref. \cite{Kamada:2019ewe}, and can potentially allow for greater parameter freedom in the Ekpyrotic contraction case we consider here.

At the end of the kination epoch the entropy density will be given by $ s=\frac{2 \pi^2}{45} g_* T_{\textrm{rh}}^3 $ with the reheating temperature no longer fixed by the bounce Hubble scale, while the lepton number density will be diluted by a factor $\frac{H_{\textrm{rh}}}{H_b} $~. Taking these factors into account, we arrive at the following prediction for the baryon asymmetry parameter, 
\begin{equation}
\frac{\eta_B^{\textrm{kin}}}{\eta_B^{obs}}\simeq \left(\frac{\epsilon}{ 350}\right)^{5/2} \left(\frac{\mathcal{N}}{1}\right) \left(\frac{N_{R-L}}{3}\right)\left(\frac{\Lambda_e}{10^{16}~\textrm{GeV}}\right)^4 \left(\frac{ |H_b|}{10^{12}~\textrm{GeV}}\right)^{2} \left(\frac{ 10^{9}~\textrm{GeV}}{T_{\textrm{rh}}}\right)~, \label{eta_kin}
\end{equation}
where $ T_{\textrm{rh}} $ has a lower limit of 100 GeV, to ensure the required equilibrium sphaleron redistribution occurs before the EWPT. 

This result can clearly replicate the observed baryon asymmetry $\frac{\eta_B^{\textrm{kin}}}{\eta_B^{obs}}\sim 1 $ for a wide range of parameter choices, due to the new freedom afforded by the easing of the reheating temperature constraint. The enhancement of the generated asymmetry compared to the inflationary scenario is evident in the dependence on the fast roll parameter. 


\section{Gravitational Leptogenesis through Chiral Gauge Field Production}
\label{Grav_lep_path2}

In this section, we investigate the generation of a non-zero $ R \tilde{R} $  by the chiral gravitational waves induced by the production of helical gauge fields. Consider a pseudoscalar field $ \varphi $ coupled to a $ U(1) $ gauge field Chern-Simons term of the following form,
\begin{equation}
\frac{\varphi}{4\Lambda}X_{\mu\nu}\tilde X^{\mu\nu}~,
\label{cs_x}
\end{equation}
where  $X_{\mu\nu}$  denotes the gauge field strength tensor of $ X_\mu $, with corresponding coupling constant $ g_{X} $, and the dual of the field strength tensor is defined as $\tilde X^{\mu\nu}= \frac{1}{2\sqrt{-g}}\epsilon^{\mu\nu\rho\sigma} X_{\rho\sigma}$. This form  of interaction can be present in low energy effective field theories associated with a Stueckelberg field 	\cite{Stueckelberg:1900zz}, or the Green-Schwarz mechanism \cite{Green:1984sg}, with corresponding UV cut-off $ \Lambda $. A coupling of this type with the inflaton or Ekpyrotic scalar and the Hypercharge Chern-Simons term has been found  to successfully  generate the observed baryon asymmetry \cite{Anber:2015yca,Jimenez:2017cdr,Domcke:2019mnd, Barrie:2020kpt}.  

The  time variation of $ \varphi $ leads to gauge field production, which carries a helical asymmetry that is dependent upon the direction of motion of $ \varphi $. Alongside the rapid production of helical gauge fields, chiral gravitational waves are also induced \cite{Cook:2011hg,Ben-Dayan:2016iks}. This has been considered as a possible route for Gravitational Leptogenesis for both abelian and non-abelian gauge fields in the context of inflation. The non-abelian $ SU(2) $ case was found to be able to lead to successful Leptogenesis \cite{Noorbala:2012fh,Maleknejad:2014wsa,Maleknejad:2016dci,Caldwell:2017chz}, but the abelian $ U(1) $ led to too small a baryon asymmetry \cite{Papageorgiou:2017yup}. The failure of the inflationary $ U(1) $ gauge field model is a result of the observational constraints on the  tensor-to-scalar ratio $ r $, and not from gauge field energy density constraints, which is the motivation for this scenario. The Ekpyrotic cosmological setting produces a blue-tilted $ r $ allowing it to circumvent this constraint, and as a result it may provide a more successful backdrop for this model. Thus, we will explore whether the abelian $ U(1) $ scenario can be successfully embedded into Ekpyrotic Cosmology.

\subsection{Gauge Field Dynamics during Ekpyrotic Contraction}

To begin, we must determine the gauge field dynamics during the contraction phase. In our model, the Ekpyrotic scalar $ \varphi $ will couple  to a $ U(1)_X $ gauge field, denoted $ X_\mu $, via the Chern-Simons coupling given in Eq. (\ref{cs_x}). The fast-rolling of $ \varphi $ induces $ \mathcal{CP} $ violating dynamics in the gauge field sector generating helical gauge fields, and associated chiral gravitational fields. Through the gravitational anomaly, a net lepton number density will be generated and subsequently redistributed to baryon number by sphalerons after reheating.  

As in the previous section, we will assume that the motion of $ \varphi $ is only negligibly affected by the Chern-Simons term and gauge field sector. For consistency  with this approximation we will require that the gauge field energy density provides a subdominant contribution to the total energy density. The action in Eq. (\ref{ekp_scalar}) and the scaling solution given in Eq. (\ref{scaling})  will be used to describe the evolution of $ \varphi $.  To obtain a positive baryon asymmetry, we require that $ \dot{\varphi}>0 $, such that the positive frequency gauge field modes are enhanced.

In a previous work, we considered the evolution of the hypercharge gauge field in an Ekpyrotic contracting background \cite{Barrie:2020kpt}. The analysis therein can be utilised here, substituting the hypercharge for the  $ U(1)_X $ gauge field. Thus, the gauge field  Lagrangian  is given by,
\begin{equation}
\frac{1}{\sqrt{-g}}{\cal L}_g=-\frac{1}{4}g^{\mu\alpha}g^{\nu\beta}X_{\mu\nu}X_{\alpha\beta} -\frac{\varphi}{4\Lambda}X_{\mu\nu}\tilde X^{\mu\nu} ~.
\label{Lagra}
\end{equation} 

The background dynamics are due to the rolling of $ \varphi $ with scale factor and Hubble rate given in Eq. (\ref{sca_fac}) and Eq. (\ref{Hub}), respectively. In conformal coordinates, the metric can be defined as $g_{\mu\nu}=a^2(\tau)\eta_{\mu\nu}$, so that the gauge field Lagrangian becomes,
\begin{equation}
{\cal L} = -\frac{1}{4}\eta^{\mu\rho}\eta^{\nu\sigma}X_{\mu\nu}X_{\rho\sigma} -\frac{\varphi}{8\Lambda}\epsilon^{\mu\nu\rho\sigma} X_{\mu\nu} X^{\rho\sigma}~.
\label{Ch3L2}
\end{equation}    

To allow analytical treatment, we will make the simplifying assumption that the back-reaction on the motion of $ \varphi $ due to the production of the  gauge field $ X_i $ is negligible. Thus, the equation of motion for the $ X $ gauge field is,
\begin{equation}
\left(\partial_{\tau}^2-\vec \bigtriangledown^2 \right) X^{i}+\frac{\varphi^\prime(\tau)}{\Lambda}\epsilon^{ijk}\partial_j X_k=0~,
\label{Ch3xfieldeom}
\end{equation}
where the gauge $X_0=\partial_i X_i=0$ has been chosen.

The dynamics of the Ekpyrotic scalar is defined by the scaling solution,
\begin{equation}
\varphi^\prime =\sqrt{\frac{2}{\epsilon}}\frac{M_p}{-\tau} ~,
\end{equation}
which upon substituting into the equation of motion for $X_{\mu}$ gives,
\begin{equation}
\left(\partial_{\tau}^2-\vec \bigtriangledown^2 \right) X^{ i}+\frac{2\kappa }{-\tau}\epsilon^{ijk}\partial_j X_k=0~,
\label{Ch3xfieldeom1}
\end{equation}
where we define the parameter,
\begin{equation} 
 \kappa =\frac{M_p}{\sqrt{2\epsilon} \Lambda}~, \label{kappa}
\end{equation} 
which is analogous to in the inflationary case, in which the instability parameter is defined as $ \xi=\sqrt{\frac{\epsilon_{\textrm{inf}}}{2}}\frac{ M_p}{ \Lambda} $ \cite{Anber:2015yca}.

  To  quantize this model, we promote the $ X $ gauge  fields to operators and assume that the boson has two  circular polarisation states,
\begin{equation}
X_i=\int\frac{d^3\vec k}{(2\pi)^{3/2}}\sum_{\alpha}\left[F_{\alpha}(\tau,k)\varepsilon_{i\alpha}\hat a^a_{\alpha} {\rm e}^{i\vec k\cdot\vec x}+
F^{*}_{\alpha}(\tau,k)\varepsilon^{*}_{i\alpha}\hat a_{\alpha}^{a\dagger}{\rm e}^{-i\vec k\cdot\vec x}
\right]~,
\label{Ch311}
\end{equation}
where $\vec \varepsilon_{\pm}$ denotes the two possible helicity states of the $X$ gauge boson ($\vec \varepsilon_{+}^{*}=\vec \varepsilon_{-}$) and the creation, $\hat a_{\alpha}^{\dagger}(\vec k)$, and annihilation, $\hat a_{\alpha}(\vec k)$, operators satisfy the canonical commutation relations,
\begin{equation}
\left[\hat a_{\alpha}(\vec k), \hat a_{\beta}^{\dagger}(\vec k')\right]=\delta_{\alpha\beta}\delta^3(\vec k-\vec k')~,
\end{equation}
and      
\begin{equation}
\hat a^a_{\alpha}(\vec k)\vert 0\rangle_{\tau}=0~,
\label{Ch314}
\end{equation}
where $\vert 0\rangle_{\tau}$ is an instantaneous vacuum state at time $\tau$.

Hence, the wave mode functions in Eq. (\ref{Ch311}) are described by the following relation,
\begin{equation}
F''_{\lambda}+\left(k^2 -\lambda  \frac{2\kappa k}{-\tau}  \right)F_{\lambda}=0~,
\label{Ch3modefunc}
\end{equation}  
where $ \lambda=\pm $ defines the helicities of the gauge field.

Solving for the mode functions $ F_{\lambda} $ in Eq. (\ref{Ch3modefunc}) gives,
\begin{equation}
F_{\lambda}=  \frac{e^{-i k \tau}}{ \sqrt{2 k}  } e^{\lambda \pi \kappa/2} U\left(i \lambda \kappa, 0, 2 i k \tau\right)~,
\label{wave_mfn}
\end{equation}
where $ U $ is a Confluent Hypergeometric functions. To derive this solution we used the Wronskian normalisation and matched to the $\mathcal{CP}$-invariant plane wave modes at $ \tau \rightarrow -\infty $, 
\begin{equation}
A_{\lambda}(\tau,k)=\frac{1}{\sqrt{2k}} e^{- ik\tau}~,
\label{BD}
\end{equation}
this wave mode equation is equivalent to that found in the inflationary setting.

We can then obtain a constraint on the model parameters such that the gauge field energy density generated by the rolling of $ \varphi $ does not come to dominate the background dynamics during the Ekpyrotic phase, that is $ 3 M_p^2 H^2 \gg \rho_{X}(\tau) $.  The energy density produced by the Chern-Simons coupling at a given $ \tau $ is approximately given by \cite{Jimenez:2017cdr},
\begin{equation} 
	3 M_p ^2H^2 \gg 1.3 \times 10^{-4}\frac{e^{2\pi \kappa}}{\kappa^3 }(\epsilon H)^4 ~,
\end{equation}
where the Hubble rate is defined in Eq. (\ref{Hub}). This relation can be rewritten as,
\begin{equation}
	\quad \frac{\epsilon^2 H}{M_p } \ll 150 \kappa^{3/2}e^{-\pi \kappa}~,
\end{equation}
which will be a key constraint on the allowed  parameter space.

To proceed to the calculation of the chiral gravitational waves, it is helpful to consider the following  approximate solution to the exponentially enhanced positive frequency wave mode equation,
\begin{align}
F_{+}(\tau,k) \simeq \frac{1}{\sqrt{2k}} \left(\frac{-k\tau}{2\kappa}\right)^{1/4} e^{\pi \kappa - 2\sqrt{-2\kappa k\tau}}~, \textrm{~~~and~~~}
F_{+}'(\tau,k)\simeq \sqrt{\frac{2k\kappa}{-\tau}} F_{+}(\tau,k)  ~,    
\label{approx}   
\end{align}
which is valid for the range $ \frac{1}{8 \kappa}<-k\tau< 2\kappa $ .  It is assumed that the exponential enhancement of the positive frequency modes in combination with the exponential suppression of the negative frequency modes, means that only the positive frequency modes need to be considered. This approximation was used in the analyses undertaken in the inflationary Gravitational Leptogenesis and Baryogenesis scenarios \cite{Anber:2015yca,Papageorgiou:2017yup}, and the calculation we follow for the gravitational waves generated by the Chern-Simons coupling between the Ekpyrotic scalar and $ U(1)_X $ gauge field \cite{Ben-Dayan:2016iks}.

\subsection{Induced Gravitational Waves and Resultant Lepton Asymmetry}
\label{grav_lep_inst}

Utilising the wave mode function solution given in Eq. (\ref{approx}) we now determine the induced chiral gravitational waves. To do so, we follow the formalism used in Ref. \cite{Ben-Dayan:2016iks}.
Once again, the gravitational wave dynamics will be determined by considering the following perturbed FRW metric,
\begin{equation}
ds^2 = a(\tau)^2(-d\tau^2 + (\delta_{ij} + h_{ij})dx_idx_j)~,
\end{equation}
where we have taken the transverse-traceless gauge for $h_{ij}$, and the scale factor is defined as $ a=(\epsilon H_b  \tau)^\frac{1}{\epsilon-1} $ . Decomposing the tensor perturbations in terms of the two possible circular polarisations,
	\begin{equation}
\hat{h}_{ij}=\frac{2}{M_p a}\int \frac{d^3k}{(2\pi)^{3/2}}e^{i \vec{k}\vec{x}}\sum_{\lambda=\pm}\Pi^*_{ij,\lambda}(\hat k)\hat Q_{\lambda}(\tau ,\vec{k})~,
\end{equation}
and thus,
\begin{equation}
\hat{h}_{\lambda}(\tau,\vec{k})\equiv \Pi_{ij,\lambda}\hat{h}_{ij}(\tau,\vec{k})=\frac{2}{M_p a}\hat{Q}_{\lambda}(\tau,\vec{k})~,
\label{h_pert}
\end{equation}
where $ \lambda=\pm $ defines the circular polarisations of the tensor perturbations. The tensor perturbation source term induced by the gauge field production is defined by,
	\begin{equation}
J_{\lambda}(\tau, \vec{k})=-\frac{a^3}{M_p }\Pi_{ij,\lambda}(\hat k)\int \frac{d^3x}{(2\pi)^{3/2}}e^{-i \vec{k}\vec{x}}\left[\hat E_i \hat E_j+\hat B_i \hat B_j\right](x^i)~,
\end{equation}
where the gauge field components are derived from the approximate solution given in Eq. (\ref{approx}), 
	\begin{align}
\hat E_i^{(\lambda)}&=-\frac{1}{a^2}\varepsilon_i^{\lambda}(\hat k)\sqrt{\frac{2 k \kappa}{-\tau}}\tilde F_{\lambda}(\tau,k)\left[\hat a_{\lambda}(\vec k)+\hat{a}_{\lambda}^{\dagger}(-\vec k)\right]~,\\
\hat B_i^{(\lambda)}&=\frac{1}{a^2}\varepsilon_i^{\lambda}(\hat k)\lambda k \tilde F_{\lambda}(\tau,k)\left[\hat a_{\lambda}(\vec k)+\hat{a}_{\lambda}^{\dagger}(-\vec k)\right]~.
\end{align}

 Evaluating the source term equation gives,
 	\begin{align}
J_{\lambda}(\tau, \vec{k})  =-&\frac{1}{M_p a} \int  \frac{d^3p}{(2\pi)^{3/2}}\sum_{\lambda'=\pm}\varepsilon_i^{(\lambda)*}(\vec k)\varepsilon_j^{(\lambda)*}(\vec k)\varepsilon_i^{\lambda'}(\vec p)\varepsilon_j^{\lambda'}(\vec k -\vec p)\left\{\frac{2 \kappa}{-\tau}\sqrt{p|\vec k-\vec p|}+p|\vec k-\vec p|\right\} \nonumber \\ &
\times\tilde X_{\lambda'}(\tau, \vec p)\tilde X_{\lambda'}(\tau,\vec k-\vec p)\left[\hat a_{\lambda'}(\vec p)+\hat{a}_{\lambda'}^{\dagger}(-\vec p)\right]
\left[\hat a_{\lambda'}(\vec k-\vec p)+\hat{a}_{\lambda'}^{\dagger}(-\vec k+\vec p)\right]~.
\end{align}
which upon substituting into Eq. (\ref{h_pert}), gives us the following relation for the  tensor perturbations,
	\begin{align}
\hat h_{\lambda}=-&\frac{2}{M_p ^2a(\tau)}\int^{\tau}d\tau'\frac{G_k(\tau,\tau')}{a(\tau')}\int \frac{d^3p}{(2\pi)^{3/2}}P_{\lambda}(\vec k, \vec p,\vec k-\vec p)
\left\{\frac{2 \kappa}{-\tau'}\sqrt{p|\vec k-\vec p|}+p|\vec k-\vec p|\right\} \nonumber\\
&\times\tilde F_{\lambda'}(\tau', \vec p)\tilde F_{\lambda'}(\tau',\vec k-\vec p)\left[\hat a_{\lambda'}(\vec p)+\hat{a}_{\lambda'}^{\dagger}(-\vec p)\right]
\left[\hat a_{\lambda'}(\vec k-\vec p)+\hat{a}_{\lambda'}^{\dagger}(-\vec k+\vec p)\right]~,
\label{gravpert}
\end{align}
where 
\begin{equation}
P_{\lambda}(\vec k, \vec p,\vec k-\vec p)=\varepsilon_i^{(\lambda)*}(\vec k)\varepsilon_i^{+}(\vec p)\varepsilon_j^{(\lambda)*}(\vec k)\varepsilon_j^{+}(\vec k -\vec p)~,
\end{equation}
and $ G_{\epsilon}(\tau, \tau') $ is the Green's function in the Ekpyrotic background. 
The relevant Green's function is given by the following, 
\begin{equation}
G_{\epsilon}(\tau, \tau')=
\frac{\Theta(\tau-\tau')\Gamma\left(\frac{1}{2}-\frac{1}{\epsilon}\right)}{2^{\frac{1}{2}-\frac{1}{\epsilon}}k}(-k \tau)^\frac{1}{\epsilon}\sqrt{-k\tau'}J_{\frac{1}{2}-\frac{1}{\epsilon}}(-k\tau')~,
\end{equation}
this Green's function is valid for both Ekpyrotic and kination epochs. In the case of kination it vanishes meaning the production of gravitational waves ceases, while when taking the limit $\epsilon \ll 1$ we obtain the function for the Ekpyrotic background,
\begin{equation}
G_{\textrm{ekp}}(\tau,\tau') \simeq\Theta(\tau-\tau')\frac{\sin(-k\tau')}{k}~.
\end{equation}

This Green's function and the approximate wave mode function in Eq. (\ref{gravpert}) can now be substituted into the tensor perturbation equation,
	\begin{align}
\hat h_{\lambda}
=-&\frac{e^{2\pi \kappa}}{M_p ^2}\int \frac{d^3p}{(2\pi)^{3/2}}P_{\lambda}(\vec k, \vec p,\vec k-\vec p) (p|\vec k-\vec p|)^{1/4} I(\tau,p,k,\kappa) \nonumber\\
& \times
\left[\hat a_{\lambda'}(\vec p)+\hat{a}_{\lambda'}^{\dagger}(-\vec p)\right]
\left[\hat a_{\lambda'}(\vec k-\vec p)+\hat{a}_{\lambda'}^{\dagger}(-\vec k+\vec p)\right]~, 
\end{align}
where
\begin{equation}
I(\tau,p,k,\kappa)=\int^{\tau}d\tau'G_{\textrm{ekp}}(\tau,\tau') 
\left(\frac{2 \kappa}{-\tau'}+\sqrt{p|\vec k-\vec p|}\right) \sqrt{\frac{-\tau'}{2\kappa}}e^{-2\sqrt{-2\kappa \tau'}\left(\sqrt{p}+\sqrt{|\vec k-\vec p|}\right)}~,
\end{equation}
and the limit $ \epsilon \ll 1 $ has been taken. 

Now that we have the above solution for the tensor perturbations, we can calculate the associated lepton number density through the following relation,
\begin{equation}
n_L= \frac{N_{R-L}}{768 \pi^4} \int dk\left( k^5 \left( \langle \hat h_+, \hat h_+\rangle-\langle \hat h_-,\hat h_-\rangle \right)-k^3 \left( \langle \hat h_+^{\prime},\hat h_+^{\prime}\rangle-\langle \hat h_-^{\prime},\hat h_-^{\prime}\rangle \right)\right)~,
\label{nL2}
\end{equation}
which is equivalent to that given in Eq. (\ref{nLden}), used in Section \ref{Grav_lep_path1}.

 The $ \langle \hat h_{\lambda},\hat h_{\lambda} \rangle $ component is given by,
	\begin{equation}
\langle \hat h_{\lambda},\hat h_{\lambda}\rangle=
\frac{2e^{4\pi \xi} }{M_p^4 }
\int \frac{d^3p}{(2\pi)^3} |P_{\lambda}(\vec k, \vec p, \vec k-\vec p)|^2 \sqrt{p|\vec k-\vec p|} I(\tau,p,k,\kappa)^2~,
\end{equation}
while for $ \langle \hat h^\prime_{\lambda}\hat, h^\prime_{\lambda} \rangle $ we have,
	\begin{equation}
\langle \hat h^\prime_{\lambda},\hat h^\prime_{\lambda}\rangle=
\frac{2e^{4\pi \xi} }{M_p^4 }
\int \frac{d^3p}{(2\pi)^3} |P_{\lambda}(\vec k, \vec p, \vec k-\vec p)|^2 \sqrt{p|\vec k-\vec p|} \left(\frac{dI}{d\tau}\right)^2~,
\end{equation}
where
	\begin{align}
\label{absP}
|P_{\lambda}(\vec k, \vec p, \vec k-\vec p)|^2=\frac{1}{16}\left(1+\lambda \frac{\vec k \cdot \vec p}{kp}\right)^2 \left(1+\lambda \frac{k^2-\vec k \cdot \vec p}{k|\vec k-\vec p|}\right)^2 \nonumber \\
=\frac{(1\pm \cos \theta)^2\left(1-q\cos \theta\pm\sqrt{1-2q \cos \theta+q^2}\right)^2}{16(1-2q \cos \theta+q^2)}~,
\end{align}
with the angle between $\vec k$ and $\vec p$ defined as $\theta$, and assuming  $\vec p=|k|\vec q$.

To determine the final lepton number density, we must analyse the parameter dependence of Eq. (\ref{nL2}) numerically after factoring out dimensional parameters. The integral is found to be dominated by the $ q= \mathcal{O}(1) $ parameter region, and it is possible to determine an approximate analytical solution of the form,
	\begin{equation}
n_L\simeq  5\cdot 10^{-12}\frac{e^{4 \pi \kappa}}{\kappa^{5}} \frac{(\epsilon H_b)^{7}}{M_p^{4}}~,
\end{equation}
where we have set $ N_{R-L}=3 $. In the inflationary case, it was shown that the integral is mostly composed of modes satisfying $ k\tau \leq 1 $, so we utilise the same scheme here such that the UV momentum  cut-off is given by $ \epsilon H_b $.   The approximation derived here is valid for the range $ 1<\kappa<4 $, but a parameter dependence similar to the inflationary case was found for $ 3<\kappa<7 $  \cite{Papageorgiou:2017yup}. This approximate equivalence between the two results is not unexpected, from the Baryogenesis mechanism considered in \cite{Jimenez:2017cdr,Barrie:2020kpt}, and the model presented in Section \ref{Grav_lep_path1}.

This allows us to calculate the final predicted baryon asymmetry parameter, assuming the usual equilibrium sphaleron redistribution and no additional entropy production after reheating. Arriving at the final predicted baryon asymmetry parameter ratio,
\begin{equation}
\frac{\eta_{B}}{\eta^{\textrm{obs}}_{B}}\simeq 0.003  \frac{e^{4 \pi \kappa}}{\kappa^{5}} \epsilon^{3/2} \left(\frac{\epsilon H_b}{M_p}\right)^{11/2}~,
\label{asym2}
\end{equation}
where we have assumed that the bounce takes place at the Hubble rate $ H_b $, followed by instantaneous reheating. Although the parameter dependence is similar to that found in the inflationary case, our result includes an enhancement due to the presence of the fast roll parameter $ \epsilon $.

\begin{figure}[t]
	\centering
	\begin{subfigure}
		\centering
		\includegraphics[width=0.4\textwidth]{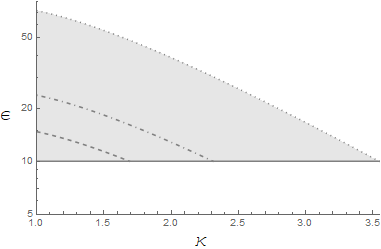}
	\end{subfigure}
	\hfill
	\begin{subfigure}
		\centering
		\includegraphics[width=0.4\textwidth]{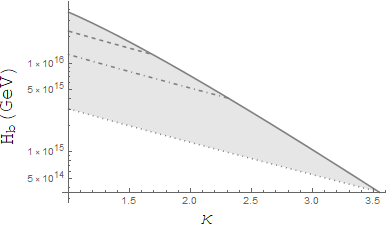}
	\end{subfigure}
	\begin{subfigure}
		\centering
		\includegraphics[width=0.4\textwidth]{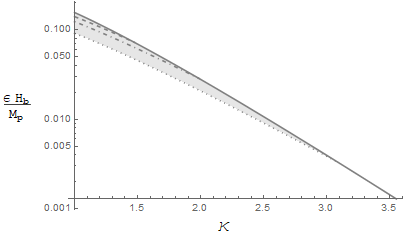}
	\end{subfigure}
	\hfill
	\begin{subfigure}
		\centering
		\includegraphics[width=0.4\textwidth]{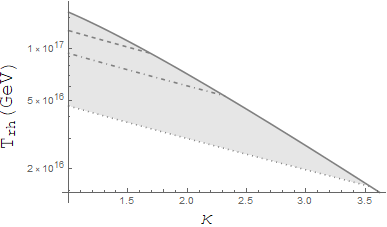}
	\end{subfigure}
	\caption{    The parameter values, (a) $ \epsilon $, (b)  $ H_b $  , (c) $ \frac{\epsilon H_b}{M_p} $, (d) $ T_{\textrm{rh}} $, as a function of $ \kappa $  that lead to successful Gravitational Leptogenesis, where a minimal value of $ \epsilon=10 $ (solid) has been chosen. Assuming instantaneous reheating, and including lines of total gauge field energy densities defined by  $ \gamma=0.1~ \textrm{(dashed)},~ 0.2 ~\textrm{(dot-dashed)},~1 ~\textrm{(dotted)} $.   }\label{constraint}
\end{figure}
To examine the parameter space for which successful Gravitational Leptogenesis can be achieved, we first define the following relation from the energy density constraint,
\begin{equation}
\frac{\epsilon H_b}{M_p}=\frac{152 \kappa^{3/2}}{e^{\pi \kappa}} \frac{\sqrt{\gamma}}{\epsilon}~,
\end{equation}
where the $ X $ gauge field energy density is defined by $ \rho_X=3 \gamma  M_p^2 H^2 $.  Upon substituting this into Eq. (\ref{asym2}) and assuming $ \eta_{B}\simeq \eta^{\textrm{obs}}_{B} $, we obtain the following requirement on the fast roll parameter, 
\begin{equation}
\epsilon\simeq 233 \frac{\kappa^{13/16}\gamma^{11/16}}{e^{3 \pi \kappa/8}}~,
\label{eps_constraint}
\end{equation}
and subsequently the dimensionless quantity,
\begin{equation}
\frac{\epsilon H_b}{M_p}=0.65\frac{\kappa^{11/16}}{e^{5 \pi \kappa/8}\gamma^{3/16}}~.
\label{dim_constraint}
\end{equation}
The bounce point Hubble rate is given by,
\begin{equation}
H_b=2.8 \cdot 10^{-3} M_p\frac{1}{\kappa^{1/8}e^{ \pi \kappa/4}\gamma^{7/8}}~,
\label{hub_constraint}
\end{equation}
with a corresponding reheating temperature of,
\begin{equation}
T_{\textrm{rh}}= 0.029 M_p\frac{1}{\kappa^{1/16}e^{\pi \kappa/8}\gamma^{7/16}}~.
\label{reh_constraint}
\end{equation}


The parameter regions subtended by the constraints defined in Eq. (\ref{eps_constraint}-\ref{reh_constraint}) are depicted in Figure \ref{constraint}, where we have assumed a lower bound on the fast roll parameter of $\epsilon\geq 10$. If we require that $ \epsilon H_b \leq 10^{16}$ GeV, as in Section 3, we obtain the  constraint $ 3.55>\kappa > 2.9$. The smallest $ \gamma $ for which successful Gravitational Leptogenesis occurs is $ \gamma \sim 0.057 $. This means we require significant gauge field energy densities within certain parameter regions, which would require consideration of possible back reaction effects.

\begin{figure}[t]
	\centering
	\includegraphics[width=0.48\textwidth]{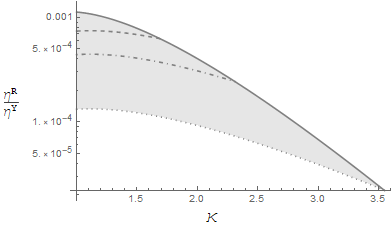}
	\caption{ The ratio of the predicted baryon asymmetry parameters for the Gravitational Leptogenesis scenario $ \eta_R $  and Hypercharge gauge field scenario $ \eta_Y $. The asymmetry parameters have been evaluated within the parameter region for which $ \eta_R \simeq \eta_{B}^{\textrm{obs}}$ .}\label{eta_ratio}
\end{figure}

It is interesting to consider whether the $ U(1) $ gauge field could be identified with the $ U(1)_Y $ of the SM \cite{Barrie:2020kpt}. In a previous work, we considered the possibility for Baryogenesis through this coupling and found a large parameter space in which successful Baryogenesis could be achieved \cite{Barrie:2020kpt}. The hypercharge mechanism presented therein, is more efficient at generating the observed baryon asymmetry than one relying on Planck suppressed gravitational couplings. Due to this, the contribution to the baryon asymmetry from the chiral gravitational waves would be expected to be negligible. In Figure \ref{eta_ratio}, we assume the parameter range that leads to successful Gravitational Leptogenesis and determine the ratio between this and that produced by the least efficient hypermagnetic field case. It can be seen that the  generated hypermagnetic field helicity asymmetry will lead to a massive overproduction of the baryon asymmetry. Thus, for successful Gravitational Leptogenesis through helical gauge fields to occur, the existence of a new $ U(1) $ gauge field is required.

\subsection{Inclusion of Kination Phase prior to Reheating}

If there is a period of kination following the bounce, it is possible to enhance the lepton number density relative to the entropy density. This is due to the differing evolution of the matter energy density ($\rho_m \propto a^{-3} $) and the scalar kinetic energy density ($\rho_m \propto a^{-6} $). In a kination expansion epoch, the energy stored in the scalar  field dilutes faster than that of the lepton number density, meaning that once reheating takes place the asymmetry parameter  has been
 enhanced. In the model presented here, we must take into account the evolution of radiation energy density generated in the $ U(1) $ gauge field during this epoch ($\rho_m \propto a^{-4} $), as reheating must occur before the radiation energy density dominates that of the scalar field. Thus, for the cases of large gauge field energy density production during the Ekpyrotic phase, we tend to require that the kinetic phase is short and subsequently provides little enhancement.

 \begin{figure}[t]
 	\centering
 	\begin{subfigure}
 		\centering
 		\includegraphics[width=0.42\textwidth]{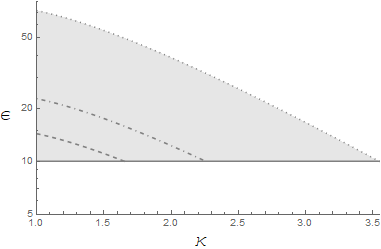}
 	\end{subfigure}
 	\hfill
 	\begin{subfigure}
 		\centering
 		\includegraphics[width=0.42\textwidth]{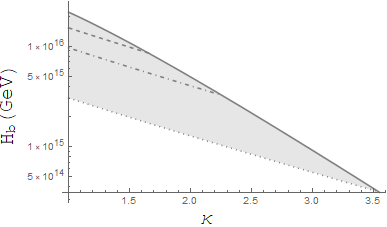}
 	\end{subfigure}
 	\begin{subfigure}
 		\centering
 		\includegraphics[width=0.42\textwidth]{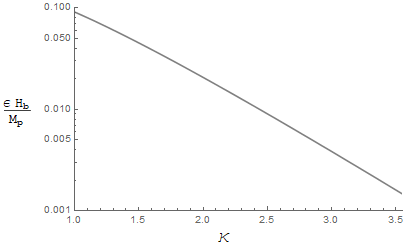}
 	\end{subfigure}
 	\hfill
 \begin{subfigure}
 	\centering
 	\includegraphics[width=0.42\textwidth]{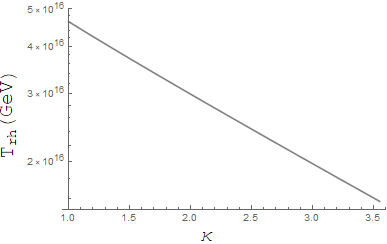}
 \end{subfigure}
 	\caption{    The parameter values, (a) $ \epsilon $, (b) $ H_b $  , (c) $ \frac{\epsilon H_b}{M_p} $, (d) $ T_{\textrm{rh}} $, as a function of $ \kappa $  that lead to successful Gravitational Leptogenesis, where a minimal value of $ \epsilon=10 $ (solid) has been chosen. Assuming a kination epoch prior to reheating, and including lines of total gauge field energy densities defined by  $ \gamma=0.04~ \textrm{(dashed)},~ 0.1 ~\textrm{(dot-dashed)},~1 ~\textrm{(dotted)} $.    }\label{constraint_kin}
 \end{figure}

  In the kination scenario, 	the predicted baryon asymmetry is given by,
 \begin{equation}
 \frac{\eta_{B}}{\eta^{\textrm{obs}}_{B}}\simeq 0.003 \frac{e^{4 \pi \kappa}}{\kappa^{5}} \frac{\epsilon^{3/2}}{\gamma^{3/4}} \left(\frac{\epsilon H_b}{M_p}\right)^{11/2}~,
 \label{asym2kin}
 \end{equation}
 where the additional $ \gamma $ factor is the maximal enhancement that can result from the inclusion of an expanding kination epoch. Analogously to the instantaneous reheating case above, we can derive the parameter relations that guarantee successful Baryogenesis, 
 \begin{equation}
 \epsilon\simeq 233 \frac{\kappa^{13/16}\sqrt{\gamma}}{e^{3 \pi \kappa/8}}~,
 \label{eps_constraint_kin}
 \end{equation}
 which gives,
 \begin{equation}
 0.091>\frac{\epsilon H_b}{M_p}=0.65\frac{\kappa^{11/16}}{e^{5 \pi \kappa/8}}>0.0015~,
 \label{dim_constraint_kin}
 \end{equation}
 and
 \begin{equation}
 H_b=2.8 \cdot 10^{-3} M_p\frac{1}{\kappa^{1/8}e^{ \pi \kappa/4}\sqrt{\gamma}}~,
 \label{hub_constraint_kin}
 \end{equation}
 from which we can calculate the expected reheating temperature range,
 \begin{equation}
 4.6 \cdot 10^{16}~\textrm{GeV}>	T_{\textrm{rh}}=6.9 \cdot 10^{16}~\textrm{GeV}~ \frac{1}{\kappa^{1/16}e^{ \pi \kappa/8}}>1.7 \cdot 10^{16}~\textrm{GeV}~.
 \label{reh_constraint_kin}
 \end{equation}

The parameter regions subtended by the constraints in Eq. (\ref{eps_constraint_kin}-\ref{reh_constraint_kin}) are depicted in Figure \ref{constraint_kin}, where we have assumed a lower bound on the fast roll parameter of $\epsilon\geq 10$. It is interesting to note that in the kination reheating case, the $ \gamma $ dependence factors out in the quantity $ \frac{\epsilon H_b}{M_p} $ and the reheating temperature. The reheating temperature coincides with $ \gamma=1 $ in the instantaneous reheating scenario, meaning lower $ T_{\textrm{rh}} $, alongside a reduction in the required maximum bounce Hubble rate for successful Gravitational Leptogenesis. If we require that $ \epsilon H_b \leq 10^{16}$ GeV, we obtain approximately the same allowed $ \kappa $ range as for the instantaneous reheating case  $ 3.55>\kappa > 2.9$. On the other hand, the smallest $ \gamma $ for which successful Gravitational Leptogenesis occurs is reduced to $ \gamma \sim 0.02 $.

\subsection{Gravitational Wave Signatures from Gauge Field Production}

The enhanced production of gauge fields is known to be able to generate unique gravitational signatures \cite{Cook:2011hg,Ben-Dayan:2016iks}, which can provide an additional avenue for observational testing of these models. To see whether there are observational consequences in our model, we can compare those sourced by the abelian gauge field generation due to the fast rolling of $ \varphi $ with those that are characteristic of Ekpyrotic Cosmologies, given in Eq. (\ref{Ekp_GW}). The gravitational waves produced by gauge field production during an Ekpyrotic contraction phase have been calculated in Ref. \cite{Ben-Dayan:2016iks}, and found to exhibit a bluer spectrum than that already predicted in Ekpyrotic Cosmology. Namely,
\begin{equation}
\mathcal{P}^{s}_{T}\simeq 3.3 \cdot 10^{-7} \frac{e^{4\pi \kappa}}{\kappa^2 }  \frac{k^3 H_b}{M_p^4 } ~,
\label{Ekp_GW22}
\end{equation}
where we have assumed $ \epsilon\gg 1 $. Comparing this to the vacuum contribution,
\begin{equation}
\mathcal{P}^{v}_{T}\simeq \frac{4 k^2}{\pi^2 M_p^2}~,
\label{Ekp_GW1}
\end{equation}
we observe the enhanced $ k $ dependence in the sourced gravitational wave spectrum.

\begin{figure}[t]
	\centering
	\begin{subfigure}
		\centering
		\includegraphics[width=0.4\textwidth]{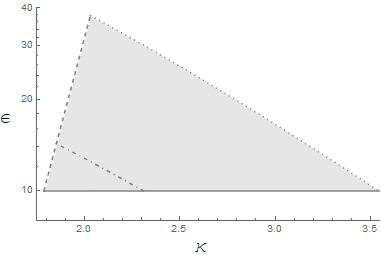}
	\end{subfigure}
	\hfill
	\begin{subfigure}
		\centering
		\includegraphics[width=0.4\textwidth]{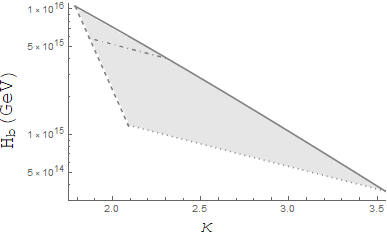}
	\end{subfigure}
	\begin{subfigure}
		\centering
		\includegraphics[width=0.4\textwidth]{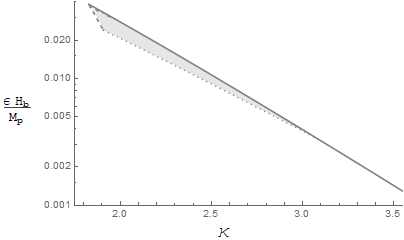}
	\end{subfigure}
	\hfill
	\begin{subfigure}
		\centering
		\includegraphics[width=0.4\textwidth]{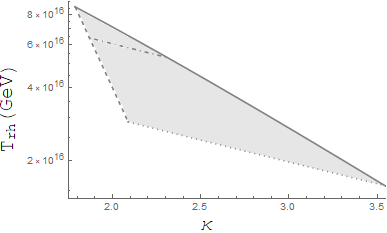}
	\end{subfigure}
	\caption{   The parameter values, (a) $ \epsilon $,  (b) $ H_b $  , (c) $ \frac{\epsilon H_b}{M_p} $, (d) $ T_{\textrm{rh}} $, as a function of $ \kappa $  that lead to successful Gravitational Leptogenesis and possible gravitational wave signals from gauge field production, where a minimal value of $ \epsilon=10 $ (solid) has been chosen. Assuming instantaneous reheating, and including lines of total gauge field energy densities defined by  $ \gamma=0.2 ~\textrm{(dot-dashed)},~1 ~\textrm{(dotted)} $.   }\label{constraint_grav}
\end{figure}

  The two components of the gravitational wave spectrum are independent, $\mathcal{P}^{\textrm{Total}}_{T}(k)=\mathcal{P}^{v}_{T}(k) + \mathcal{P}^{s}_{T}(k) $, so it is possible to determine the frequencies for which the gravitational waves sourced from  gauge field production become dominant. The frequency range of observational interest, for which $ \mathcal{P}^{s}_{T}(k) \ge \mathcal{P}^{v}_{T}(k) $ and successful Gravitational Leptogenesis is achieved, is given by,
\begin{equation}
24 \textrm{ GHz} \left(\frac{\epsilon}{15}\right)^{\frac{4}{11}} \left(\frac{\kappa}{3}\right)^{\frac{16}{11}}  e^{\frac{4 \pi}{11}(3-\kappa)}>f> 1.7 \textrm{ MHz} \left(\frac{\epsilon}{15}\right)^{\frac{21}{11}} \left(\frac{\kappa}{3}\right)^{\frac{7}{11}}  e^{\frac{32 \pi}{11}(3-\kappa)} ~,
\end{equation}
where the upper frequency limit is derived from the cut-off $ k\tau_b=2\kappa $, above which the effects of gauge field amplification are suppressed. This range is for the instantaneous reheating result found in Eq. \ref{asym2}. Therefore, the gauge field production can generate features in the high frequency region of the gravitational wave spectrum, which  if possible to probe in the future, could provide important information about the details of the Ekpyrotic mechanism and the origin of the matter-antimatter asymmetry. In Figure \ref{constraint_grav}, we show the parameter regions for which successful Gravitational Leptogenesis coincides observationally relevant gravitational waves from gauge field production.

\begin{figure}[t]
	\centering
	\begin{subfigure}
		\centering
		\includegraphics[width=0.4\textwidth]{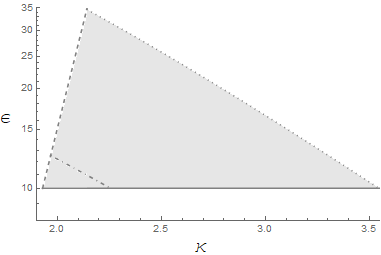}
	\end{subfigure}
	\hfill
	\begin{subfigure}
		\centering
		\includegraphics[width=0.4\textwidth]{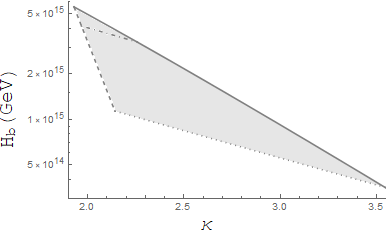}
	\end{subfigure}
	\begin{subfigure}
		\centering
		\includegraphics[width=0.4\textwidth]{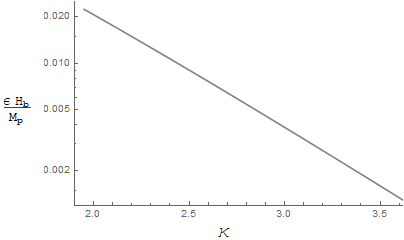}
	\end{subfigure}
	\hfill
	\begin{subfigure}
		\centering
		\includegraphics[width=0.4\textwidth]{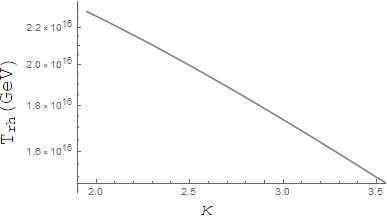}
	\end{subfigure}
	\caption{    The parameter values, (a) $ \epsilon $,  (b) $ H_b $  , (c) $ \frac{\epsilon H_b}{M_p} $, (d) $ T_{\textrm{rh}} $, as a function of $ \kappa $  that lead to successful Gravitational Leptogenesis and possible gravitational wave signals from gauge field production, where a minimal value of $ \epsilon=10 $ (solid) has been chosen. Assuming a kination epoch prior to reheating, and including lines of total gauge field energy densities defined by  $ \gamma=0.1 ~\textrm{(dot-dashed)},~1 ~\textrm{(dotted)} $.  }\label{constraint_grav_kin}
\end{figure}

Similarly to the instantaneous reheating case discussed above, we can derive the observationally relevant frequency range for the kination scenario. Utilising Eq. (\ref{asym2kin}), assuming successful Gravitational Leptogenesis, we obtain the following frequency range,
\begin{equation}
24 \textrm{ GHz} \sqrt{\frac{\epsilon}{15}}  \left(\frac{\kappa}{3}\right)^{\frac{43}{32}} e^{\frac{5 \pi}{16}(3-\kappa)}>f> 2 \textrm{ MHz} \left(\frac{\epsilon}{15}\right)^{\frac{3}{2}} \left(\frac{\kappa}{3}\right)^{\frac{31}{32}}  e^{\frac{49 \pi}{16}(3-\kappa)} ~,
\end{equation}
where once again the upper frequency limit is derived from the cut-off $ k\tau_b=2\kappa $. As an example, consider the largest allowed $ \kappa \sim 3.55 $, for which we find the following frequency range of interest $ 14 \textrm{ GHz} >f> 6 \textrm{ kHz} $~. Future experiments will be required to probe the high frequency gravitational waves generated in this mechanism.

\section{Conclusion}
\label{conc}
 An Ekpyrotic contraction epoch is a unique venue for investigating mechanisms for Baryogenesis, with inherent advantages for models typically considered in an inflationary context. This is alongside its ability to provide solutions to the known cosmological problems as well as those associated with inflationary cosmology. We have investigated the possibility of Gravitational Leptogenesis taking place during an Ekpyrotic contraction phase.  In this mechanism, the generation of chiral gravitational waves during the cosmological evolution leads to the production of a net lepton asymmetry through the gravitational anomaly.  Two possible ways to construct this mechanism were explored, namely, coupling the fast-rolling Ekpyrotic scalar to the gravitational Chern-Simons term, and to an abelian gauge field Chern-Simons interaction.

In the gravitational Chern-Simons interaction case, we found that successful Leptogenesis can occur in both the instantaneous reheating and kination scenarios. This is in contrast to the inflationary scenario, in which an extended period of kination is necessary.  The allowed parameter space in the instantaneous reheating case is limited by the requirement that no graviton ghost modes exist and the validity of the WKB approximation, with only a small allowed parameter region. In Ekpyrotic Cosmology, periods of kinetic contraction and expansion are expected, and allowing for the possibility of an extended kination expansion can have significant implications for Leptogenesis. In applying this to Gravitational Leptogenesis, we find ample parameter space for which successful Baryogenesis can occur. An additional benefit of the Ekpyrotic context over the inflationary setting is the ability to consistently define the Bunch-Davies vacuum state at early times.

Another way to generate the chiral gravitational waves necessary for Gravitational Leptogenesis, is through enhanced helical gauge field production. By coupling the Ekpyrotic scalar to the Chern-Simons term of an abelian gauge field, we have found that a significant lepton number asymmetry can be produced. A large region of parameter space for successful Gravitational Leptogenesis was found for both the instantaneous reheating and kination expansion scenarios. This is in contrast to the inflationary scenario which is unable to explain the observed matter-antimatter asymmetry in both cases. This scenario also predicts high frequency gravitational wave signatures for much of the parameter space, which may provide an additional path for experimental verification.

 Future theoretical and experimental developments will be integral to probing this mechanism for generating the observed matter-antimatter asymmetry.  Improved precision in the measurement of non-gaussianities and a detailed analysis of the predictions in this Ekpyrotic scenario, may play a key role in constraining the allowed parameter space. As discussed in Section \ref{EKP}, the non-gaussianities produced in Ekpyrotic Cosmology can be consistent with current observational measurements but significant uncertainties still exist.  This observable is dependent upon the details of the Ekpyrotic scalar sector, the period around the bounce point, and possible non-gaussianities from the Chern-Simons couplings. Thus, a dedicated analysis must be completed, which shall be undertaken in future work. Gravitational wave observables will also provide a key method for exploring the allowed parameter space. The discovery of a  tensor-to-scalar ratio in near future CMB experiments could place significant constraints on the Ekpyrotic scenario \cite{Hazumi:2019lys}, while gravitational wave observations in the high frequency regime may detect the chiral gravitational waves produced in the gauge field production mechanism for Gravitational Leptogenesis. This result motivates further investigation into the rich phenomenology of the Ekpyrotic scenario and its possible implications for our understanding of the early universe.

\section*{Acknowledgements}
This work was supported  by IBS under the project code, IBS-R018-D1, and by the World Premier International Research Center Initiative (WPI), MEXT, Japan.

\end{document}